\documentclass[prd,reprint,showpacs,showkeys]{revtex4-1}
\usepackage{amsfonts}
\usepackage{amssymb}
\usepackage{amsmath}
\usepackage{graphicx}
\usepackage[font={footnotesize,it}]{caption}

\begin{document}

\title{Thin-shell wormholes in 2+1-dimensional Einstein-Scalar Theory}
\author{S. Habib Mazharimousavi}
\email{habib.mazhari@emu.edu.tr}
\author{Z. Amirabi,}
\email{zahra.amirabi@emu.edu.tr}
\author{M. Halilsoy}
\email{mustafa.halilsoy@emu.edu.tr}
\affiliation{Department of Physics, Eastern Mediterranean University, Gazima\~{g}usa,
Turkey. }
\date{\today }

\begin{abstract}
We present an infinite class of one-parameter scalar field extensions to the
BTZ black hole in 2+1-dimensions. By virtue of the scalar charge the
thin-shell wormhole supported by a linear fluid at the throat becomes stable
against linear perturbations. More interestingly, we provide an example of
thin-shell wormhole which is strictly stable in the sense that it is
confined in between two classically intransmissible potential barriers.
\end{abstract}

\maketitle

\section{Introduction}

Since the discovery of Ba\~{n}ados, Teitelboim and Zanelli (BTZ) black hole
in $2+1-$dimensions, there has been a vast literature in the same context 
\cite{1,2,3,4,5,6}. Expectedly there has been a race to derive rival metrics
to BTZ by employing new physical sources other than the cosmological
constant \cite{7,8,9,10,11,12,13,14,15}. As the problem amounts to introduce
new physical sources it has become part of the classical field theory in
curved spacetime. Just to mention a few, we note that they range from linear
Maxwell \cite{16} to non-linear electrodynamics \cite{14} as well as scalar
fields \cite{8,9,10}, dilatons \cite{7,17,18,19,20,21,22,23,24,25,26},
scalar multiplets \cite{27}, Born-Infeld \cite{28}, Brans-Dicke extensions 
\cite{29}, etc.. In particular, recently we have obtained classes of scalar
multiplet fields acting as sources of gravity in both $2+1$ and $3+1-$%
dimensions \cite{27}. This method amounts to consider a set of scalar
multiplets $\left\{ \phi ^{a}\right\} $ with the index '$a$' belonging to an
internal gauge group, specifically the group $O\left( 2\right) $ for $2+1$
and $O\left( 3\right) $ for $3+1-$dimensions. However, upon deriving the
field equations we had the freedom to reduce the problem to circular /
spherical symmetry by suppressing the angular dependence and concentrating
only on the modulus of $\left\{ \phi ^{a}\right\} $. The advantage in such a
reduction ansatz is to pave the way towards technically solvable system of
equations. The scope of the model can further be extended by adding a
self-interacting potential. From physical considerations scalar multiplets
are familiar field theoretical objects whose experimental construction may
be feasible in a foreseeable future. The special case of a scalar multiplet
is provided by a scalar singlet in which there is no internal gauge group .
Our model considered herein is a massless scalar field singlet with an
exponential potential of scalar coupled to gravity in $2+1-$dimensions \cite%
{27}. Our system admits exact solutions with two parameters and can be
considered yet another, one-parametric extension of the BTZ solution. After
substitution of the solution for the scalar field the potential takes the
form $V\left( r\right) =\frac{\text{const.}}{r^{\alpha ^{2}}},$ where $%
\alpha $ is the crucial parameter in our model ranging over $0\leq \alpha
<\infty ,$ unless stated otherwise. Note also that the solution for $-\infty
<\alpha <0$ can easily be mapped to the former case so that we make our
choice of $\alpha $ in the range $0\leq \alpha <\infty .$ It can be
anticipated easily that for $\alpha =0$ the resulting black hole solution
reduces to the BTZ solution. We make use of the parameter $\alpha $, which
can be phrased as "scalar charge" together with the cosmological constant to
establish a large class of Einstein-scalar solutions. Once this is
accomplished the principal aim in the present paper is to employ this class
of solutions to build thin-shell wormholes \cite%
{30,31,32,33,34,35,36,37,38,39,40,41,42,43,44,45,46,47,48,49,50,51,52,53,54,55,56}%
. The role of the scalar field in such a construction is interesting:
assuming a linear gas / fluid equation of state at the throat of the
thin-shell wormhole allows us to make stable such wormholes. By stability it
is meant throughout the paper, against small linear perturbations to
preserve its identity through harmonic oscillations. Initial velocities at
the throat of the perturbed fields may be considered in various ranges
bounded only by the speed of light. These are called velocity dependent
perturbations and they play important role in stability / instability
checking analysis. When these are all taken into consideration we are able
to show the existence of stability regions for our wormhole all by virtue of
the role played by the scalar charge. Besides stability against linear
perturbations we construct also thin-shell wormhole confined in between high
potential barriers so that it remains forever stable in its potential well.

Organization of the paper is as follows. In Sec. II we present the
Einstein-scalar field equations with the exponential potential and integrate
them completely. Construction and dynamical analysis of the wormhole is
investigated thoroughly in Sec. III. In Sec. IV we complete the paper with
our concluding remarks.

\section{$2+1-$dimensional Einstein-scalar solutions}

Before we start this section we comment that, recently a different approach
has been employed by two of us to find similar results in Ref. \cite{27}.
The action in $2+1-$dimensional spherically symmetric, static theory of
gravity coupled to a scalar field is given by ($8\pi G=1$) 
\begin{equation}
S=\int d^{3}x\sqrt{-g}\left( R-\frac{1}{2}\partial _{\mu }\psi \partial
^{\mu }\psi -\lambda e^{-\alpha \psi }\right)
\end{equation}%
in which $R$ is the Ricci scalar, $\psi =\psi \left( r\right) $ is the
scalar field, $\alpha $ and $\lambda $ are coupling constants.

The line element is chosen to be%
\begin{equation}
ds^{2}=-A\left( r\right) dt^{2}+\frac{dr^{2}}{B(r)}+r^{2}d\theta ^{2}
\end{equation}%
in which $A\left( r\right) $ and $B\left( r\right) $ are unknown functions
of radial coordinate $r$ to be found from the field equations. Explicitly
the Lagrangian density becomes%
\begin{equation}
\mathcal{L}=\sqrt{-g}R-\frac{r}{2}\sqrt{AB}\left( \psi ^{\prime }\right)
^{2}-\lambda r\sqrt{\frac{A}{B}}e^{-\alpha \psi }
\end{equation}%
where a prime denotes $\frac{d}{dr}$. The scalar field equation is obtained
as%
\begin{equation}
\psi ^{\prime \prime }+\psi ^{\prime }\left( \ln \left( r\sqrt{AB}\right)
\right) ^{\prime }+\frac{\lambda \alpha }{B}e^{-\alpha \psi }=0.
\end{equation}%
Einstein field equations can be derived directly by variation of the action
in terms of $g^{\mu \nu }$ and they are given by%
\begin{equation}
G_{\mu \nu }=T_{\mu \nu }
\end{equation}%
where $T_{\mu \nu }$ is the energy momentum tensor given by 
\begin{equation}
T_{\mu \nu }=\frac{1}{2}\left( \partial _{\mu }\psi \partial _{\nu }\psi -%
\frac{1}{2}\partial _{\gamma }\psi \partial ^{\gamma }\psi g_{\mu \nu
}\right) -\frac{1}{2}\lambda e^{-\alpha \psi }g_{\mu \nu }.
\end{equation}%
Due to $\psi \left( r\right) $ which is a function of $r$ the explicit form
of the energy momentum tensor components are calculated as 
\begin{equation}
T_{tt}=\left( \frac{1}{2}\left( \psi ^{\prime }\right) ^{2}B+\lambda
e^{-\alpha \psi }\right) \frac{A}{2},
\end{equation}%
\begin{equation}
T_{rr}=\frac{1}{2B}\left( \frac{1}{2}\left( \psi ^{\prime }\right)
^{2}B-\lambda e^{-\alpha \psi }\right)
\end{equation}%
and%
\begin{equation}
T_{\theta \theta }=-\frac{r^{2}}{2}\left( \frac{1}{2}\left( \psi ^{\prime
}\right) ^{2}B+\lambda e^{-\alpha \psi }\right) .
\end{equation}%
Finally the explicit form of the Einstein's equations are given by%
\begin{equation}
\frac{B^{\prime }}{r}+\frac{1}{2}\left( \psi ^{\prime }\right) ^{2}B+\lambda
e^{-\alpha \psi }=0,
\end{equation}%
\begin{equation}
\frac{BA^{\prime }}{r}-\frac{A}{B}\left( \frac{1}{2}\left( \psi ^{\prime
}\right) ^{2}B-\lambda e^{-\alpha \psi }\right) =0
\end{equation}%
and%
\begin{multline}
\frac{2ABA^{\prime \prime }-BA^{\prime 2}+A^{\prime }B^{\prime }A}{2}+ \\
A^{2}\left( \frac{1}{2}\left( \psi ^{\prime }\right) ^{2}B+\lambda
e^{-\alpha \psi }\right) =0
\end{multline}%
which together with (4) admit the following solutions%
\begin{equation}
\psi \left( r\right) =\alpha \ln r
\end{equation}%
\begin{equation}
A\left( r\right) =\left\{ 
\begin{array}{cc}
C_{2}\left( r^{2}+\frac{\left( \alpha ^{2}-4\right) C_{1}}{2\lambda }r^{%
\frac{\alpha ^{2}}{2}}\right) & \alpha \neq 2 \\ 
C_{2}r^{2}\left( C_{1}-\lambda \ln r\right) & \alpha =2%
\end{array}%
\right.
\end{equation}%
and%
\begin{equation}
B\left( r\right) =\left\{ 
\begin{array}{cc}
\frac{2\lambda }{\alpha ^{2}-4}r^{-\alpha ^{2}}\left( r^{2}+\frac{\left(
\alpha ^{2}-4\right) C_{1}}{2\lambda }r^{\frac{\alpha ^{2}}{2}}\right) & 
\alpha \neq 2 \\ 
\frac{1}{r^{2}}\left( C_{1}-\lambda \ln r\right) & \alpha =2%
\end{array}%
\right.
\end{equation}%
in which $C_{1}$ and $C_{2}$ are two integration constants.

Let's add that with $\alpha =0$ one finds $\psi =0$ and the action reduces to%
\begin{equation}
S=\int d^{3}x\sqrt{-g}\left( R-\lambda \right)
\end{equation}%
in which $\lambda $ plays the role of a cosmological constant. The field
equations admit the BTZ solution with \cite{1,2,3}%
\begin{equation}
A=B=-M_{BTZ}-\frac{\lambda }{2}r^{2}.
\end{equation}%
Therefore for $\alpha \neq 2$ we set $C_{1}=-M_{BTZ}$ and $C_{2}=\frac{%
2\lambda }{\alpha ^{2}-4}$ such that the $\alpha \rightarrow 0$ limit works
correctly. Hence our metric functions for $0\leq \alpha <2$ can be written as%
\begin{equation}
A\left( r\right) =\frac{2\lambda }{\alpha ^{2}-4}\left( r^{2}-\frac{\left(
\alpha ^{2}-4\right) M_{BTZ}}{2\lambda }r^{\frac{\alpha ^{2}}{2}}\right)
\end{equation}%
and%
\begin{equation}
B\left( r\right) =\frac{2\lambda }{\alpha ^{2}-4}r^{-\alpha ^{2}}\left(
r^{2}-\frac{\left( \alpha ^{2}-4\right) M_{BTZ}}{2\lambda }r^{\frac{\alpha
^{2}}{2}}\right)
\end{equation}%
with the potential 
\begin{equation}
V\left( r\right) =-\frac{\lambda }{r^{\alpha ^{2}}}.
\end{equation}%
For the case $\alpha =2,$ from (14) and (15) one can redefine the time and
absorb the constant $C_{2}$ which means that we can set it to one.

Depending on the values of the free parameters, the general solution may
admit black hole or naked singular spacetimes. But at any cost the solution
is not asymptotically flat. Therefore to find the mass of the central object
one can use the definition of Brown and York (BY) mass \cite{57,58}. Before
we apply this formalism we would like to comment that other methods exist in
literature which use the Noether's theorem in order to find the conserved
quantities including mass in a non-asymptotically flat spacetime. For $2+1-$%
dimensional BTZ solution we suggest to consult \cite{59,60} while for a
wider study and comparison between these methods with the BY prescription
one may look at \cite{61}. Since our aim in the present study is to
investigate Einstein Scalar thin-shell wormholes we delete the alternative
analysis for the conserved quantities to a future study.

In accordance with the BY mass in a $D-$dimensional spherically symmetric
spacetime with the line element 
\begin{equation}
ds^{2}=-F\left( r\right) ^{2}dt^{2}+\frac{dr^{2}}{G\left( r\right) ^{2}}%
+r^{2}d\Omega _{D-2}^{2}
\end{equation}%
the BY mass is given by%
\begin{multline}
M_{BY}= \\
\lim_{r_{b}\rightarrow \infty }\frac{D-2}{2}r_{b}^{D-3}F\left( r_{b}\right)
\left( G_{ref}\left( r_{b}\right) -G\left( r_{b}\right) \right)
\end{multline}%
where $G_{ref}\left( r_{b}\right) $ is the reference function and $r_{b}$
stands for the radius of the boundary surface. Using (21) for the black hole
solution with $0\leq \alpha <2$ we obtain $M_{BY}=\frac{M_{BTZ}}{4}$ and the
line element can be written as%
\begin{multline}
ds^{2}=-\frac{4}{4-\alpha ^{2}}\left( \frac{r^{2}}{\ell ^{2}}-\left(
4-\alpha ^{2}\right) M_{BY}r^{\frac{\alpha ^{2}}{2}}\right) dt^{2}+ \\
\frac{r^{\alpha ^{2}}dr^{2}}{\frac{4}{4-\alpha ^{2}}\left( \frac{r^{2}}{\ell
^{2}}-\left( 4-\alpha ^{2}\right) M_{BY}r^{\frac{\alpha ^{2}}{2}}\right) }%
+r^{2}d\theta ^{2}.
\end{multline}%
in which $-\frac{\lambda }{2}=\frac{1}{\ell ^{2}}.$ This solution includes
the BTZ limit when $\alpha \rightarrow 0.$ This form of line element is
valid only when $0\leq \alpha <2.$ For $\alpha >2$ the solution does not
admit any known limit and therefore one has to consider the solution in its
original form given by (14) and (15). In the rest of this work we impose $%
0\leq \alpha <2$ and the line element which we shall refer frequently is
given by (23). Furthermore, $\alpha <0$ can be mapped to $\alpha >0$ and
therefore we exclude it.

Before we finish this section we add that the field equations, in the limit
of $\lambda \rightarrow 0$ admit 
\begin{equation}
A\left( r\right) =\zeta r^{\frac{\alpha ^{2}}{2}},
\end{equation}%
\begin{equation}
B\left( r\right) =\frac{1}{A\left( r\right) }
\end{equation}%
with $V\left( r\right) =0$ in which $\zeta $ is a new integration constant.
Hence the line element becomes%
\begin{equation}
ds^{2}=\zeta r^{\frac{\alpha ^{2}}{2}}\left( -dt^{2}+dr^{2}\right)
+r^{2}d\theta ^{2}
\end{equation}%
which is the solution of Einstein-self interacting scalar field in $2+1-$%
dimensions.

\section{Thin-shell wormholes}

Let's consider the $2+1-$dimensional spherically symmetric static bulk
spacetime, $\mathcal{M}$, given by (23) which can be written more
appropriately as%
\begin{multline}
ds^{2}=-\frac{4}{4-\alpha ^{2}}\frac{r^{2}}{\ell ^{2}}\left( 1-\left( \frac{%
r_{+}}{r}\right) ^{\frac{4-\alpha ^{2}}{2}}\right) dt^{2} \\
+\frac{r^{\alpha ^{2}}dr^{2}}{\frac{4}{4-\alpha ^{2}}\frac{r^{2}}{\ell ^{2}}%
\left( 1-\left( \frac{r_{+}}{r}\right) ^{\frac{4-\alpha ^{2}}{2}}\right) }%
+r^{2}d\theta ^{2}.
\end{multline}%
where $r_{+}$ is the radius of the horizon which is given by 
\begin{equation}
r_{+}=\left( \ell ^{2}\left( 4-\alpha ^{2}\right) M_{BY}\right) ^{\frac{2}{%
4-\alpha ^{2}}},
\end{equation}%
with $0\leq \alpha <2$. We consider a timelike hypersurface $\Sigma $
defined by%
\begin{equation}
\digamma =r-a=0
\end{equation}%
in which $a>r_{+}.$ Next, we make two copies from the bulk located outside $%
\Sigma $, i.e., $r>a$ (let's call them $\mathcal{M}^{\left( 1,2\right) }$)
and we paste them at the location of $\Sigma .$ The resulting manifold i.e., 
$\mathcal{M}^{\left( 1\right) }\cup \mathcal{M}^{\left( 2\right) }$ is a
complete manifold with a throat located at the location of $\Sigma .$ For
the smooth matching at the throat one has to apply the Israel junction
conditions \cite{62,63,64} which are given by, i) the induced metric tensor
on either side of the throat must be identical i.e., $h_{ij}^{\left(
1\right) }=h_{ij}^{\left( 2\right) }$ and ii) the discontinuity of the
extrinsic curvature must satisfy 
\begin{equation}
k_{i}^{j}-k\delta _{i}^{j}=-8\pi GS_{i}^{j}
\end{equation}%
in which $k_{i}^{j}=K_{i}^{j\left( 2\right) }-K_{i}^{j\left( 1\right) }$, $%
k= $ $k_{i}^{i}$ and $S_{i}^{j}=diag\left( -\sigma ,p\right) $ is the energy
momentum tensor on the shell $\Sigma .$ Herein $K_{ij}^{\left( 1,2\right) }$
is the extrinsic curvature tensor on the either sides of the throat defined
by%
\begin{equation}
K_{ij}^{\left( 1,2\right) }=-n_{\gamma }^{\left( 1,2\right) }\left( \frac{%
\partial ^{2}x^{\gamma }}{\partial \xi ^{i}\partial \xi ^{j}}+\Gamma
_{\alpha \beta }^{\gamma }\frac{\partial x^{\alpha }}{\partial \xi ^{i}}%
\frac{\partial x^{\beta }}{\partial \xi ^{j}}\right)
\end{equation}%
where $x^{\alpha }=\left\{ t,r,\theta \right\} $ and $\xi ^{i}=\left\{ \tau
,\theta \right\} $ are the coordinate systems on the bulk and the throat,
respectively. The first condition is satisfied by setting%
\begin{equation}
h_{ij}^{\left( 1\right) }=h_{ij}^{\left( 2\right) }=h_{ij}=diag\left(
-1,a\left( \tau \right) ^{2}\right)
\end{equation}%
such that the line element on the shell is given by%
\begin{equation}
ds_{\Sigma }^{2}=-d\tau ^{2}+a\left( \tau \right) ^{2}d\theta ^{2}
\end{equation}%
in which $\tau $ is the proper time on the shell. Let's add also that in
general the coordinates of either sides of the throat are related as $%
r^{\left( 1\right) }=r^{\left( 2\right) }=a\left( \tau \right) $, $\theta
^{\left( 1\right) }=\theta ^{\left( 2\right) }=\theta $ and 
\begin{equation}
\left( \frac{dt^{\left( 1,2\right) }}{d\tau }\right) ^{2}=\left. \frac{1}{%
A^{\left( 1,2\right) }}\left( \frac{1}{B^{\left( 1,2\right) }}\left( \frac{%
da\left( \tau \right) }{d\tau }\right) ^{2}+1\right) \right\vert _{\Sigma }
\end{equation}%
but due to the identical metric in either sides of the throat i.e., $%
A^{\left( 1,2\right) }=A$ and $B^{\left( 1,2\right) }=B$ they become equal
i.e.,%
\begin{multline}
\left( \frac{dt^{\left( 1\right) }}{d\tau }\right) ^{2}=\left( \frac{%
dt^{\left( 2\right) }}{d\tau }\right) ^{2}= \\
\left. \frac{1}{A}\left( \frac{1}{B}\left( \frac{da\left( \tau \right) }{%
d\tau }\right) ^{2}+1\right) \right\vert _{\Sigma }
\end{multline}%
in which 
\begin{equation}
A=\frac{4}{4-\alpha ^{2}}\frac{r^{2}}{\ell ^{2}}\left( 1-\left( \frac{r_{+}}{%
r}\right) ^{\frac{4-\alpha ^{2}}{2}}\right) \text{ }
\end{equation}%
and 
\begin{equation}
B=r^{-\alpha ^{2}}A.
\end{equation}%
In (31) $n_{\gamma }^{\left( 1,2\right) }$ are the normal vectors on both
sides of the throat defined as%
\begin{equation}
n_{\gamma }^{\left( 1,2\right) }=\pm \left( \frac{\partial \digamma }{%
\partial x^{\alpha }}\frac{\partial \digamma }{\partial x^{\beta }}g^{\alpha
\beta }\right) ^{-\frac{1}{2}}\frac{\partial \digamma }{\partial x^{\gamma }}
\end{equation}%
in which based on (31) the $+$ and $-$ refer to $n_{\gamma }^{\left(
2\right) }$ and $n_{\gamma }^{\left( 1\right) },$ respectively, and
explicitly they are given by%
\begin{equation}
n_{\gamma }^{\left( 1,2\right) }=\pm \left( -\sqrt{\frac{A}{B}}\dot{a},\frac{%
\sqrt{B+\dot{a}^{2}}}{B},0\right) _{\Sigma }.
\end{equation}%
Herein a dot implies the derivative with respect to the proper time $\tau .$
Following $n_{\gamma }^{\left( 1,2\right) }$ we obtain%
\begin{equation}
k_{\tau }^{\tau }=\frac{A^{\prime }\sqrt{B+\dot{a}^{2}}}{A}+\frac{2B\ddot{a}%
-B^{\prime }\dot{a}^{2}}{B\sqrt{B+\dot{a}^{2}}},
\end{equation}%
and%
\begin{equation}
k_{\theta }^{\theta }=\frac{2\sqrt{B+\dot{a}^{2}}}{a}
\end{equation}%
which yield%
\begin{equation}
\sigma =-\frac{\sqrt{B+\dot{a}^{2}}}{4\pi Ga}
\end{equation}%
and%
\begin{equation}
p=\frac{A^{\prime }\sqrt{B+\dot{a}^{2}}}{8\pi GA}+\frac{2B\ddot{a}-B^{\prime
}\dot{a}^{2}}{8\pi GB\sqrt{B+\dot{a}^{2}}}
\end{equation}%
in which a prime stands for the derivative with respect to $r$ and all
expressions are at $r=a.$ Explicitly one can show that the energy
conservation implies%
\begin{equation}
\frac{d\left( \mathcal{A}\sigma \right) }{d\tau }+p\frac{d\mathcal{A}}{d\tau 
}=\Xi
\end{equation}%
in which $\mathcal{A}=2\pi a$ is the area of the throat and 
\begin{equation}
\Xi =\frac{\left( BA^{\prime }-AB^{\prime }\right) \sqrt{B+\dot{a}^{2}}}{4GAB%
}\dot{a}.
\end{equation}%
If we assume that there exists an equilibrium radius $a=a_{0}$ where the
throat of the wormhole is at rest with no acceleration, i.e., $\dot{a}=\ddot{%
a}=0,$ the energy density and lateral pressure become static. At such
equilibrium condition one obtains%
\begin{equation}
\sigma _{0}=-\frac{\sqrt{B_{0}}}{4\pi Ga_{0}}
\end{equation}%
and%
\begin{equation}
p_{0}=\frac{A_{0}^{\prime }\sqrt{B_{0}}}{8\pi GA_{0}}
\end{equation}%
in which a sub$-0$ stands for the equilibrium state. Therefore if the radius
of the throat is set to be $a=a_{0}$ we may consider the wormhole to be at
equilibrium but it may not be a stable state. Therefore we perturb the
throat which means we apply a force and change the radius of the throat. If
after the perturbation the throat tends to go back to its initial
equilibrium point we shall define the wormhole to be stable, otherwise
unstable. To see if the throat goes back to its equilibrium point or not,
one has to write an equation of motion for the throat. This equation is
extracted from the dynamic expression of $\sigma $ given by Eq. (42). Hence,
by rearranging the terms in (42) we find a one-dimensional classical motion
of the form%
\begin{equation}
\dot{a}^{2}+V\left( a\right) =0
\end{equation}%
where 
\begin{equation}
V\left( a\right) =B-\left( 4\pi Ga\sigma \right) ^{2}.
\end{equation}%
To go further one has to identify $\sigma $ as a function of $a.$ This can
be achieved by considering an equation of state (EoS) for the fluid
presented on the shell. Here in this study we consider a linear EoS of the
form 
\begin{equation}
p-p_{0}=w\left( \sigma -\sigma _{0}\right)
\end{equation}%
in which $w$ is a constant. EoS (50) and the energy conservation (44) can be
combined to find an explicit expression for $\sigma $. Our analytic
calculation yields%
\begin{multline}
\sigma =\frac{\left( 2+\alpha ^{2}\right) \sigma _{0}+2p_{0}}{2w+2+\alpha
^{2}}\left( \frac{a_{0}}{a}\right) ^{w+1+\frac{\alpha ^{2}}{2}} \\
+\frac{2w\sigma _{0}-2p_{0}}{2w+2+\alpha ^{2}}.
\end{multline}%
Upon (51), the potential function becomes%
\begin{multline}
V\left( a\right) =B-\left( 4\pi Ga\right) ^{2}\times \\
\left( \frac{\left( 2+\alpha ^{2}\right) \sigma _{0}+2p_{0}}{2w+2+\alpha ^{2}%
}\left( \frac{a_{0}}{a}\right) ^{w+1+\frac{\alpha ^{2}}{2}}+\frac{2w\sigma
_{0}-2p_{0}}{2w+2+\alpha ^{2}}\right) ^{2}.
\end{multline}%
%
%
%
%
%
%
%
%
%
%
%
%
%
%
%
%
%
%
%
%
%
%
%
%
%
\begin{figure}[tbp]
\includegraphics[width=70mm,scale=0.7]{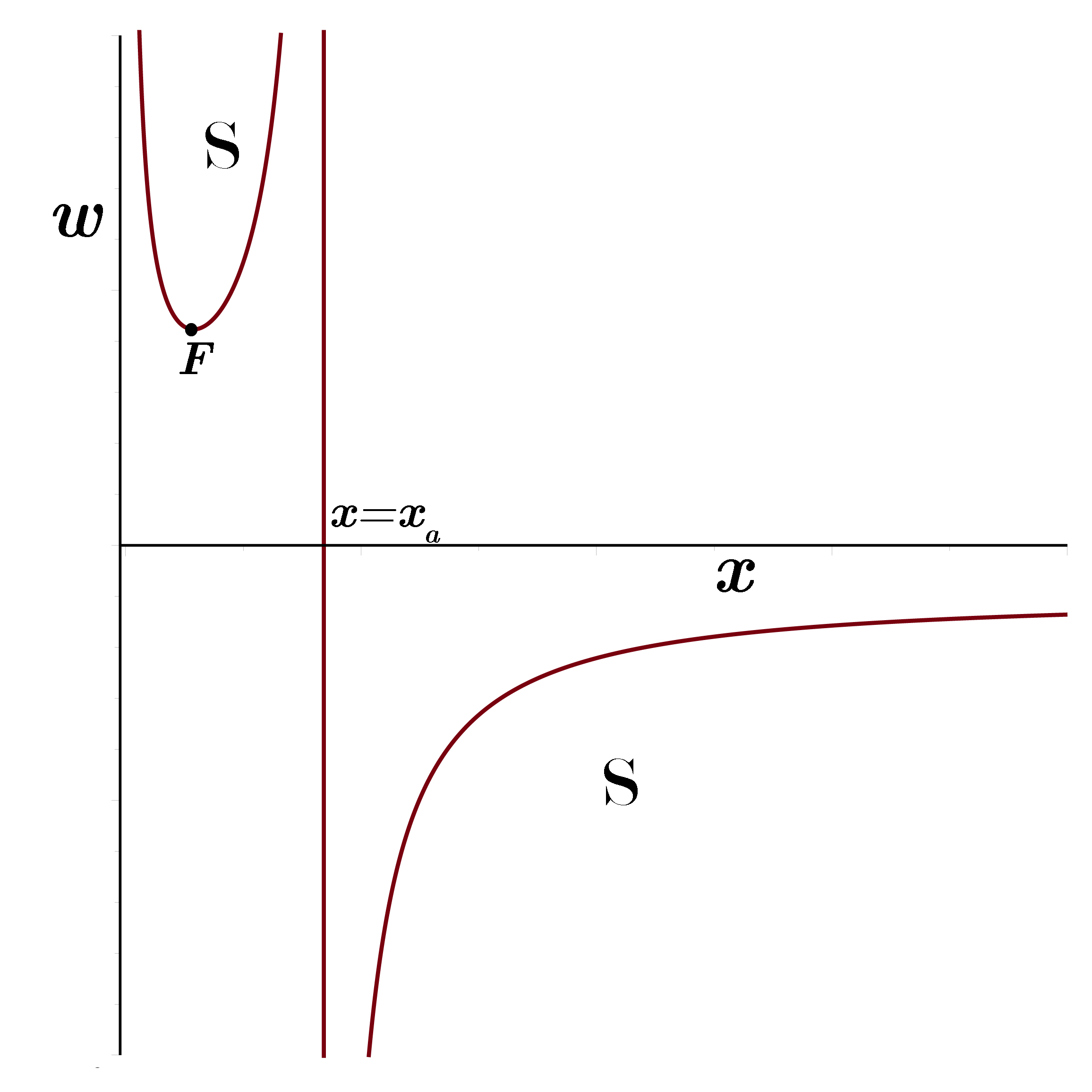}
\caption{The general behaviour of $V_{0}^{\prime \prime }$ versus $x=\frac{%
a_{0}}{r_{+}}$ and $w$ for an orbitrary $\protect\alpha ^{2}<4.$ The parts
labeled with $S$ shows the stable regions.}
\end{figure}

\begin{figure}[tbp]
\includegraphics[width=70mm,scale=0.7]{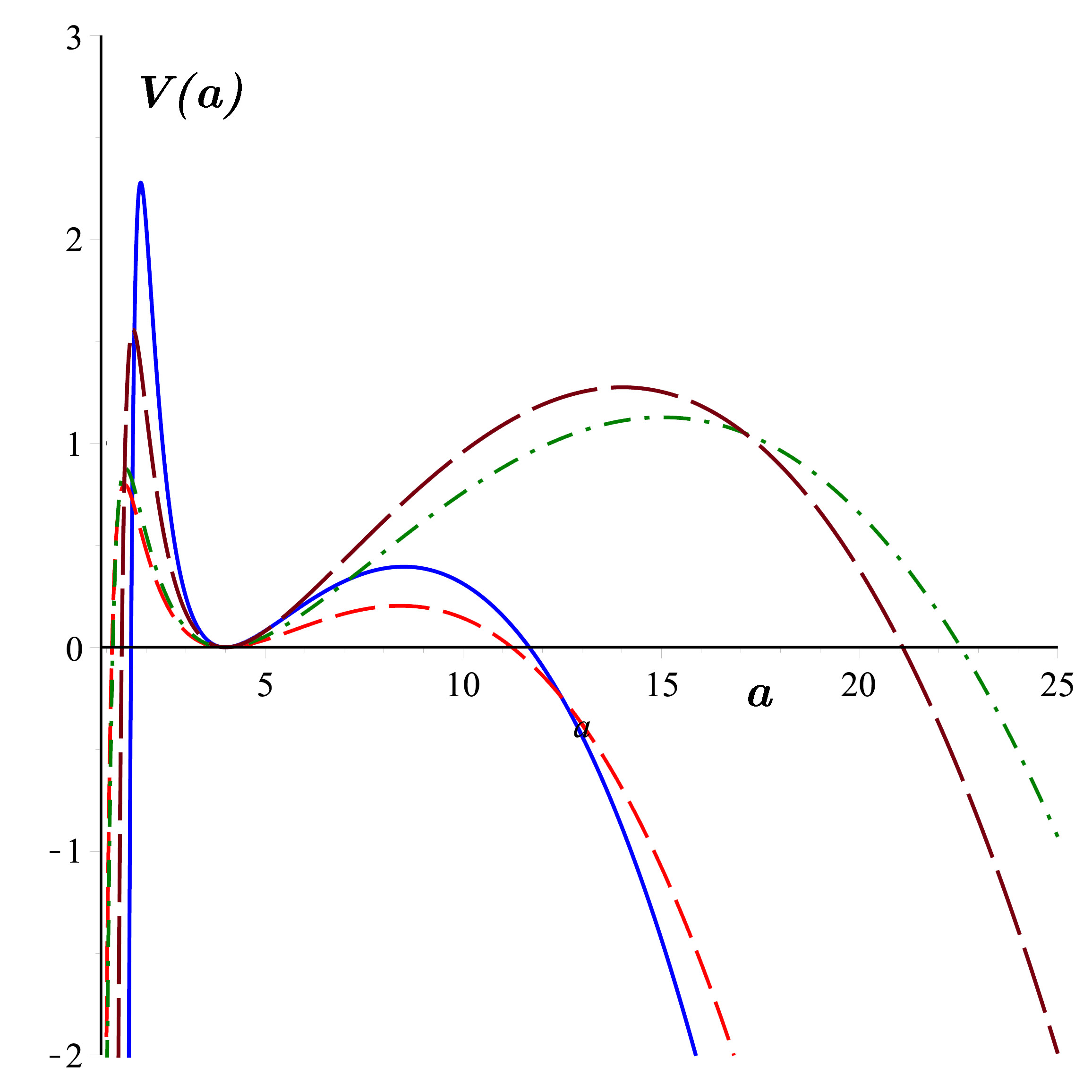}
\caption{Potential $V\left( a\right) $ versus $a$ for common $r_{+}=1$ and $%
\ell ^{2}=1$ but individually: blue / solid $w=5$, $\protect\alpha =0.2,$
red / short-dash $w=2.7$, $\protect\alpha =0.17,$ green / dash-dot $w=2.7$, $%
\protect\alpha =0.137$ and brown / long-dash $w=2.7$, $\protect\alpha %
=0.155. $ }
\end{figure}

One can explicitly show that $V\left( a\right) $ is zero when $a=a_{0}.$
Also its first derivative vanishes at the equilibrium radius but the second
derivative is obtained to be%
\begin{multline}
V_{0}^{\prime \prime }=\frac{1}{2\left( 4-\alpha ^{2}\right) \ell
^{2}a_{0}^{\alpha ^{2}}\left( 1-\left( \frac{r_{+}}{a_{0}}\right) ^{\frac{%
4-\alpha ^{2}}{2}}\right) }\times \\
\left[ 4\left( \left( 3w+5\right) \alpha ^{2}+4w-4\right) \left( \frac{r_{+}%
}{a_{0}}\right) ^{\frac{4-\alpha ^{2}}{2}}\right. - \\
\left. \left( \alpha ^{2}+4w\right) \left( \alpha ^{2}+4\right) \left( \frac{%
r_{+}}{a_{0}}\right) ^{4-\alpha ^{2}}-8\alpha ^{2}\left( 1+w\right) \right] .
\end{multline}%
To have the wormhole locally stable, $V_{0}^{\prime \prime }$ must be
positive in the vicinity of $a=a_{0}.$ Before that we note that, $r_{+}$ and 
$\ell ^{2}$ play as scale parameters. In Fig. 1 we present the general
behavior of $V_{0}^{\prime \prime }$ in terms of $w$ and $x=\frac{a_{0}}{%
r_{+}}$ but $0\leq \alpha <2$. In this figure the regions labeled by $S$ are
the stable part of the parameters. Also the point $F$ is located at 
\begin{equation}
x_{F}=\left( \frac{2\alpha ^{2}}{3\alpha ^{2}-4}-\frac{2\left( 4-\alpha
^{2}\right) \alpha }{\left( 3\alpha ^{2}-4\right) \sqrt{4+\alpha ^{2}}}%
\right) ^{\frac{-2}{4-\alpha ^{2}}}
\end{equation}%
with%
\begin{equation}
w_{F}=\frac{\left( \alpha ^{4}-7\alpha ^{2}+4\right) \sqrt{4+\alpha ^{2}}%
+\alpha ^{3}\left( 4+\alpha ^{2}\right) }{\left( 5\alpha ^{2}+4\right) \sqrt{%
4+\alpha ^{2}}-4\alpha \left( 4+\alpha ^{2}\right) }.
\end{equation}

%
\begin{figure}[tbp]
\includegraphics[width=70mm,scale=0.7]{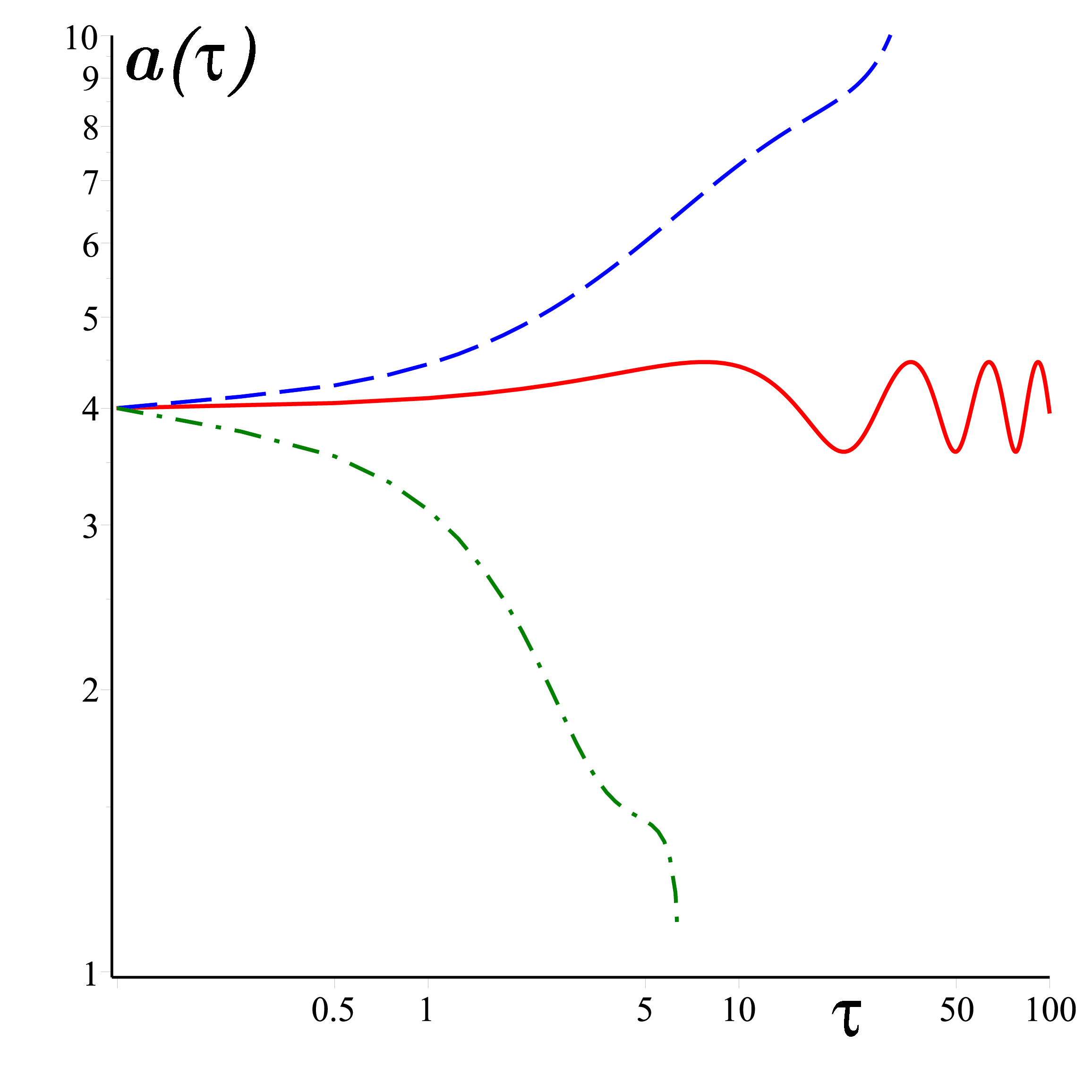}
\caption{The motion of the throat under the potential shown in Fig. 2
(red/short-dash) with three different possibilities / initial velocity. The
axis are logarithmic and dash, solid and dash-dot represent evaporation,
oscillation and collapse of the thin-shell wormhole with initial velocities $%
\left. \frac{da}{d\protect\tau }\right\vert _{\protect\tau =0}=$ $0.46$, $%
0.10$ and $-0.895$ respectively. }
\end{figure}


In addition to that the vertical asymptote is given by $x=x_{a}$ where%
\begin{equation}
x_{a}=\left( \frac{4+\alpha ^{2}}{2\alpha ^{2}}\right) ^{\frac{2}{4-\alpha
^{2}}}.
\end{equation}%
This, however, does not mean that the wormhole is stable against any kind of
perturbation as a matter of fact, it is valid only for small perturbation.
To find the exact motion of the throat after a general perturbation one
should solve the general equation of motion either analytically or
numerically. In Fig. 2 we plot the potential $V\left( a\right) $ versus $a$
for certain values of parameters which are given in its caption. We observe
that for small perturbation $a=a_{0}=4$ is a local stable equilibrium point
for all different cases. As a function of proper time the throat radius $%
a\left( \tau \right) $ is plotted in Fig. (3). It is observed that this
provides a stable, i.e. oscillatory radius wormhole within certain range of
initial velocities. Fig. (2) justifies explicitly the stability property
since the radius is confined in between two high potential barriers.

\section{Conclusion}

In Einstein's general relativity wormholes / thin-shell wormholes have two
characteristic features that often attract criticism. The first is the
occurrence of negative energy density which, with the exception of some
alternative works \cite{65} is almost taken for granted while the second
concerns about the stability of such structures. Our aim in this paper was
to consider Einstein's gravity solutions with a scalar charge and
investigate whether stable wormholes can be established. To this end we
presented first an infinite class of solutions with well-known limits.
Remarkably it is found that scalar charge ($\alpha $) gives rise to
stability regions for the underlying thin-shell wormholes. Small linear
perturbation within certain limits of initial velocity is shown to keep the
thin-shell wormhole stable. Arbitrary perturbations, expectedly, either
collapses or evaporates the wormhole. Not only through perturbations but
finely tuned Einstein-Scalar thin-shell wormholes also can be established
which are eternally stable in their deep potential well. It is our strong
belief that similar results hold also in $3+1-$dimensions, however, the
technical details may not be as straightforward as in the present $2+1-$%
dimensional case.

\end{document}